\documentstyle[12pt]{article}
\title{\nopagebreak
\begin{flushright}
\tenrm UCTP-101-96
\end{flushright}\vskip .7in
\large \bf Dimensionless Coupling of Superstrings
\\ to Supersymmetric Gauge Theories and
\\ Scale Invariant Superstring Actions}
\author{\large M. Awada\thanks{e-mail address:
moustafa@physunc.phy.uc.edu} and F. Mansouri\thanks{e-mail
address: mansouri@uc.edu} \\ \it \small 
Physics Department,
University of Cincinnati, Cincinnati, OH 45221}
\date{}

\begin{document}
\baselineskip=14pt
\maketitle

\begin{abstract}
We construct new superstring actions which are distinguished from
the standard superstrings by being space-time scale invariant. Like
standard superstrings, they are also reparametrization invariant,
space-time supersymmetric, and invariant under local scale
transformations of the world sheet. We discuss two possible
scenarios in which these actions could play a significant role, in
particular one which involves their coupling to supersymmetric
gauge theories.
\end{abstract}
\pagebreak

It is well known that string models were the outgrowth of the
attempts to formulate a theory of strong interactions. Early work
on these models made it clear that standard string theories are
not well suited for this purpose. In particular, the string based
scattering amplitudes at high energy and fixed scattering angle
fall off exponentially with energy, whereas experiments in the deep
inelastic region indicate that they should be power behaved. This
was, in fact, one of the primary reasons for reinterpreting strings
as theories incorporating all interactions in the hope that the
connection to QCD would emerge with further study. 

The power law behaviour of the scattering amplitudes suggests that
a string theory
consistent with QCD at short distances must have a phase which is
characterized by long range order
i.e., by absence of a scale. To this end, Polyakov [1] considered
modifying
the 
Nambu action by a renormalizable scale invariant term involving the
square of the curvature of the world sheet, thus arriving at the
theory of rigid strings.  
In the large N approximation, where N is the dimension of space-
time, this model does not undergo a phase
transition. One way to induce such a phase transition [2] is to
couple
the rigid string to the long range Kalb-Ramond
field [3]. It has been shown that to leading and subleading order
the resulting theory does indeed undergo a phase transition to a
region of long range order [2].

One of the aims of the present work is to provide a framework
within which such an analysis can be carried out for standard
superstrig theories. As mentioned above, the two main ingredients
of such a framework are a scale invariant addition to the
superstring action and the coupling of a long range interaction
to the superstrings characterized by a dimensionless parameter.
Once such a space-time scale invariant superstring action is
constructed, it can also be cosidered as a superstring theory in
its own right. As will be seen below, it may provide a dynamical
basis for understanding the origin of the string tension. So, we
will construct
such superstring actions without reference to a particular
application. Not having an a priori knowledge of the structure of
a space-time scale invariant superstring action, it turns out to be
more convenient to first construct its coupling to the long range
field.
As will be seen below, the regularization of this interaction will
lead us directly to the sought after superstring action.  

A prime candidate for the long range interaction is a gauge field.
Although it is possible to construct interactions between strings
and gauge fields which are characterized by a dimensionful coupling
constant [4], it can be easily seen that there are no
such dimensionless interaction terms with at most two derivatives
between strings and 
non-supersymmetric gauge theories. The only exception is 
the parity violating expression
$$  ie\int d\sigma F^*\  .\eqno{(1)}$$
In (1) $d\sigma$ is the surface measure, and $F^*$ is
the dual of the gauge field strength.
Therefore, it is of interest to explore whether supersymmetric
gauge 
theories can couple to a superstring theory with a dimensionless
coupling strength. In the interest of clarity, in this work we
confine ourselves to the abelian case. The non-abelian version will
be discussed elsewhere [5].

In a recent work [6], we showed that it is possible to extend the
notion of Wilson loop to supersymmetric gauge theories. This was
accomplished by making use of the Stoke's theorem, and its non-
abelian version, to express the Wilson loop in terms of the field
strength on a two-suface with a boundary and then seek its
supersymmetric generalization. The result was expressed in terms of
chiral superfields and chiral currents on the two-surface. The
expression for
the chiral current necessary for this supersymmetrization turned
out to be not the most general one that could be constructed on the
two surface. Its genralization led us to the construction of a new
"stringy" observable which depends on the metric and which has no
analogue in non-supersymmetric gauge theories. It is also
characterized by a dimensionless coupling constant. 

To write down and study the properties of this stringy observable,
we follow the notation and conventions of reference [6] and use the
standard two component superspace formalism [7]. Thus the
components of the supersymmetric vielbein are given by 
$$v_{a}^{\alpha{\dot \alpha}} =\partial_{a}x^{\alpha{\dot
\alpha}}(\xi)
-{i\over 2}(\theta^{\alpha}(\xi)\partial_{a}\theta^{{\dot
\alpha}}(\xi) 
+\theta^{{\dot \alpha}}(\xi)\partial_{a}\theta^{\alpha}(\xi))$$
$$v_{a}^{\alpha} =\partial_{a}\theta^{\alpha}(\xi)\eqno{(2)}$$
$$v_{a}^{{\dot \alpha}} =\partial_{a}\theta^{{\dot \alpha}}(\xi).$$
They are invariant under global space-time supersymmetry 
transformation rules defined
$$\delta x^{\alpha{\dot \alpha}}(\xi) = {i\over
2}(\epsilon^{\alpha}
\theta^{{\dot \alpha}}(\xi) +\epsilon^{{\dot
\alpha}}\theta^{\alpha}(\xi))$$
$$\delta \theta^{\alpha}(\xi) = \epsilon^{\alpha}\eqno{(3)}$$
$$\delta \theta^{{\dot \alpha}}(\xi) = \epsilon^{{\dot \alpha}}.$$
The requirement that, e.g, in four dimensions, the coordinates
$\theta$ 
satisfy the Majorana condition demands that $\epsilon$ be a
Majorana. 
Thus we define: 
$$C_{ab}^{\alpha} =v_{a}^{\alpha{\dot \alpha}}v_{b {\dot \alpha}}
\eqno{(4)}$$
$$C_{ab}^{{\dot \alpha}} =v_{a}^{\alpha{\dot \alpha}}v_{b
\alpha}.$$

The field content of the supersymmetric Maxwell theory can be
expressed in terms of the chiral superfield
$W_{\alpha}(x,\theta)$ and its conjugate. They satisfy the
chirality conditions:
$$D_{\alpha}W_{{\dot \beta}} = D_{{\dot \alpha}}W_{\beta } =
0\eqno{(5)}$$
$$D^{{\dot \alpha}}W_{{\dot \alpha}} = D^{\alpha}W_{\alpha}.$$
The $W$'s are determined in terms of an unconstrained vector 
superfield $V$:
$$W_{\alpha} = {-i\over 2}{\bar D}^2D_{\alpha}V~~;~~ W_{{\dot
\alpha}} =
{i\over 2}D^2D_{{\dot \alpha}}V\eqno{(6)}$$
which are solutions of (5). The $W's$ are invariant under the gauge
transformation
$$\delta V = i({\bar \Lambda} - \Lambda)\eqno{(7)}$$
where $\Lambda$ (${\bar \Lambda}$) is a chiral (anti-chiral)
parameter superfield.
The component expansion of $V$ and $W_{\alpha}$ in the Wess-Zumino
gauge are
respectively,
$$V = (0,0,0,0, A_{\alpha{\dot \alpha}},\psi_{\alpha},\psi_{{\dot
\alpha}}, D)
\eqno{(8 a)}$$
$$W_{\alpha} = \psi_{\alpha} - \theta^{\beta}f_{\alpha\beta} 
-i\theta_{\alpha}D
+{i\over 2}\theta^{2}\partial_{\alpha{\dot \alpha}}\psi^{{\dot
\alpha}}
\eqno{(8 b)}$$
where $\psi$ is the superpartner of the gauge field $A_{\alpha{\dot
\alpha}}$, 
$f_{\alpha\beta} = {1\over 2}\partial_{(\alpha{\dot
\alpha}}A_{\beta)}^{{\dot
\alpha}}$ is the Maxwell's field strength and D is an 
auxiliary field.  An important property of the $W_{\alpha}$ 
($W_{{\dot \alpha}}$) is that it is invariant under the chiral part
(anti-chiral part) of the supersymmetry transformations of the
component
fields:
$$\delta_{\epsilon^{\alpha}}W_{\alpha} = 
\delta_{\epsilon^{{\dot\alpha}}}W_{{\dot \alpha}} = 0\  .\eqno{(9
a)}$$
$$\delta A^{\alpha{\dot \alpha}} = i(\epsilon^{\alpha}\psi^{\dot
\alpha}
+\epsilon^{{\dot\alpha}}\psi^{\alpha})$$
$$\delta \psi_{\alpha} = \epsilon^{\beta}f_{\alpha\beta}+
i\epsilon_{\alpha}D\eqno{(9 b)}$$
$$\delta D ={1\over 2}\partial_{\alpha{\dot\alpha}}
(\epsilon^{\alpha}\psi^{\dot \alpha}
-\epsilon^{{\dot\alpha}}\psi^{\alpha})$$ 

With these preliminaries, we can now write down the stringy
interaction obtained in reference [6] :
$$S_{int.} = \kappa\int_{\Sigma}d^2\xi \sqrt{-h} h^{ab}
C_{ab}^{\alpha}(\xi)W_{\alpha}(x(\xi),\theta(\xi)) +
h.c\eqno{(10)}$$
where $h^{ab}$ is the metric on the two surface, h is its
determinant, and
$C_{ab}^{\alpha}(\xi)$ are the components of a manifestly
supersymmetric invariant
spinor 
tensor given in (4).
The quantities
$W_{\alpha}(x(\xi),\theta(\xi))$ are the supersymmetric
invariant abelian chiral superfields given above, 
and h.c denotes hermitian conjugation.  The interaction (10) is
also
invariant under the local scale transformation 
$h_{ab}\rightarrow \Lambda(\xi)h_{ab}$ of the world sheet metric.

We know of no way of constructing this interaction term in the
absence of supersymmetry. It is space-time supersymmetric, gauge
invariant,  
reparemetrization invariant, and locally scale invariant on the
world sheet. It is also characterized by a new 
coupling constant $\kappa$ which is, at least classically,
different from the gauge
coupling $e$. In four space-time dimensions, $e$ is dimensionless.
So is $\kappa$ . in space-time dimensions three, six, and ten, in
which supersymmetric gauge theories exist also, $\kappa$ is not
dimensionless but has the same dimension as $e$ . It is, however,
possible to modify the action (10) so that $\kappa$ remains
dimensionless in any of these dimesions. 

In analogy with the expression for the supersymmetric Wilson loop,
we can now define a new supersymmetric and gauge invariant
"stringy" observable 
$$\Psi(\Sigma) = e^{S_{int.}}\eqno{(11)}$$ 
If we take the surface $\Sigma$ to have a boundary, a closed loop
C, then
the correlation functions of $\Psi(\Sigma)$ might be useful for
describing loops formed by pair
creation 
and annihilation of superparticles, particularly in strongly
coupled super QED. 
As mentioned above, the superstring-like observable can be
expressed in terms
of chiral currents on the surface.  Define
$${\cal J}^{\alpha}(z) = \kappa\int_{\Sigma(C)} d^2\xi 
{\cal Q}^{\alpha}(\xi)\delta^{6}(z-z(\xi))\eqno{(12 a)}$$
where $\delta^{6}(z-z(\xi))$ is the chiral delta function which 
in the
chiral representation takes the form $=\delta^{4}(x-x(\xi))(\theta
-\theta(\xi))^2$, 
and 
$$ {\cal Q}^{\alpha}(\xi) = \sqrt{-h} h^{ab}C_{ab}^{\alpha}(\xi)
\  .\eqno{(12 b)}$$
The action (10) can then be rewritten as [6] :
$$S_{int} =\int d^6z ({\cal J}^{\alpha}W_{\alpha} + h.c)\ 
.\eqno{(13)}$$
In this form, it is manifestly supersymmetric, and gauge invariant.
 
From equation $(12b)$ it is clear why the stringy observable has no
counter part in non-supersymmetric gauge theories. In the absence
of supersymmetry, $C_{ab}^{\alpha}=0$, and we loose
the superstring-gauge field interaction (10). 

Next, we compute the expectation value, $<\Psi(C)>$, of the
stringy observable.
Since the theory is abelian one can easly show that [8]:
$$ <\Psi(\Sigma)> = e^{-{1\over 4}\int d^6z \int d^6z' ({\cal
J}^{\alpha}(z) ({\cal J}^{\beta}(z')<W_{\alpha}(z)W_{\beta}(z')> +
h.c}\  .\eqno{(14)}$$
The average in (14) is taken with respect to the Boltzmann factor
of the
super-Maxwell action
$$ S_{Super~Maxwell} = \int d^6z W^{\alpha}(z)W_{\alpha}(z)
\eqno{(15)}$$
Using the expression of $W_{\alpha}$ in (8b) one finds:
$$<W_{\alpha}(z)W_{\beta}(z')> = {1\over
2}\delta^{6}(z-z')\epsilon_{\alpha\beta}\eqno{(16)}$$
Inserting this into (14) and doing one of the z integrations we
obtain
$$ S_{string} = -{\kappa^2\over 8}\int_{\Sigma}d^2\xi
\int_{\Sigma}d^2\xi'
\delta^{4}(x(\xi)-x(\xi'))(\theta(\xi) -\theta(\xi'))^2 {\cal
Q}^{\alpha}(\xi)
{\cal Q}_{\alpha}(\xi') + h.c\  .\eqno{(17)}$$
Clearly (17) is divergent and requries regularization.  We will
regularize
the delta function by making the replacement
$$\delta^{4}(x(\xi)-x(\xi'))\rightarrow 
\delta_{\epsilon}^{4}(x(\xi)-x(\xi')) = {1\over \pi^2\epsilon^4}
e^{{(x(\xi)-x(\xi'))^2\over \epsilon^2}}\eqno{(18)}$$
and in the end taking the limit $\epsilon \rightarrow 0$.  This
regularization method can be
interpreted
geometrically by the manifold splitting regularization method where
one
displaces $\Sigma$ in the first measure infinitesimally away that
in the 
second measure along some unit normal $n^{\mu}(\xi)$.  
Thus we define $\Sigma_{\epsilon}$ to have coordinates 
$y^{\mu} = x^{\mu} + \epsilon n^{\mu}$ where $x^{\mu}$ is
the coordinate on $\Sigma$.  Inserting (18) into (17) and taking
the limit
$\epsilon \rightarrow 0$ we obtain the following action:
$$S = -{\kappa^2\over 8\pi}\int_{\Sigma}d^2\xi\sqrt{-h}
({h\over G})^{1\over 2} G^{ab}K_{ab}\eqno{(19)}$$
where $G^{ab}$ is the inverse of the induced metric $G_{ab}$ 
on the world sheet:
$$G_{ab}= v_{a}^{\mu}v_{b}^{\mu}\eta_{\mu\nu}\eqno{(20)}$$
$$K_{ab} = K^{\alpha}K_{\alpha}g_{ab} + h.c\  .\eqno{(21)}$$
$$ K^{\alpha} = h^{ab}C_{ab}^{\alpha}(\xi),~~~g_{ab}
=v_{a}^{\alpha}
v_{b{\alpha}}$$
It is straight forward to show that
$$ K_{ab} = -2 K_{a} K_{b}\eqno{(22 a)}$$
where
$$ K_{a} = K^{\alpha}v_{a{\alpha}}\  .\eqno{(22 b)}$$
Further manpulations makes the action takes the simple form
$$ S = {\kappa^2\over 4\pi}\int_{\Sigma}d^2\xi\sqrt{-h}
({h\over G})^{1\over 2}(G^{ab}- G^{cd}h_{cd} h^{ab})t_{a}t_{b}
\eqno{(23 a)}$$
where
$$t_{a} ={1\over 2}v_{a}^{\alpha{\dot \alpha}}e_{\alpha{\dot
\alpha}}
\  .\eqno{(23 b)}$$
$$e^{\alpha{\dot \alpha}} = \epsilon^{ab}\partial_{a}v_{b}
^{\alpha{\dot \alpha}}$$
The quantity $\epsilon$ is the covariant antisymmetric tensor on
the world sheet. To put our
action in a more familiar form, we resort to the 4-component
notation, with $\mu=0,...D-1$ being the space-time index, and
$\alpha
=1,...4$ the spinor index.  We have 
$$S = {\kappa^2\over 16\pi}\int_{\Sigma}d^2\xi\sqrt{-h}
({h\over G})^{1\over 2}(G^{ab} -G^{cd}h_{cd} h^{ab})v_{a}^{\mu}
v_{b}^{\nu}e_{\mu}e_{\nu}\eqno{(24 a)}$$
where
$$v_{a}^{\mu} =\partial_{a}x^{\mu}(\xi)
-i{\bar
\theta}^{\alpha}(\xi)\Gamma^{\mu}\partial_{a}\theta_{\alpha}(\xi) 
\eqno{(24 b)}$$
$$e^{\mu} = \epsilon^{ab}\partial_{a}v_{b}^{\mu}$$
The space-time vectors $e^{\mu}$ are supersymmetric and are of mass
dimension. The supersymmetric action (24a) is
invariant under the two dimensional general coordinate
transformations, space-time supersymmetry transformations,
global space-time scale transformations, and local world sheet
scale
transformations given by :
$$ h_{ab}\rightarrow \Lambda(\xi)h_{ab},~~~
e_{\mu}\rightarrow \Lambda^{-1}(\xi)e_{\mu}\  .\eqno{(25)}$$
Except for space-time scale invariance, these invariances are
shared with the Green-Schwarz action [9]. So, the distinguishing
feature of our superstring action is its invariance under space-
time scale transformations. 

The action (24a) can be further simplified.  Define the
set of supersymmetric, locally scale invariant vectors:
$$\sigma^{\mu} = {\varepsilon^{ab}\over (G)^{1\over 2}}
\partial_{a}v_{b}^{\mu}\eqno{(26)}$$  
where $\varepsilon$ is the numerical antisymmetric tensor that
transforms
as a density ($\varepsilon^{ab} =\sqrt{-h}\epsilon^{ab}$).  Then
the 
action
(24a) becomes
$$S = {\kappa^{2}\over 16\pi}\int_{\Sigma}d^2\xi\sqrt{G}
(G^{ab} -G^{cd}h_{cd} h^{ab})v_{a}^{\mu}
v_{b}^{\nu}\sigma_{\mu}\sigma_{\nu}\  .\eqno{(27)}$$
The above action is expressed in a second order formalism, and it
is more desirable to express it in its first order form. Since we
have established all the invariances of the action $(27)$, in
particular its space-time scale invariance, we can simply ask if
there are any actions consistent with these symmetries. We find
that there are only two such actions. One is given by
$$S_{0} = {\kappa_{0}^2\over 16\pi}\int_{\Sigma}d^2\xi\sqrt{-h} 
h^{ab}v_{a}^{\mu}v_{b}^{\nu}\sigma_{\mu}\sigma_{\nu}\ 
.\eqno{(28)}$$
It is the first order form of the action (27).  To write down the
other action we first define the following
composite field:
$$\Phi = \sigma^{\mu}\sigma_{\mu}\   .\eqno{(29)}$$
Then we have
$$S_{1} = {\kappa_{1}^2\over 4\pi}\int_{\Sigma}d^2\xi\sqrt{-h} 
h^{ab}G_{ab}\Phi\  .\eqno{(30)}$$
We note in passing that an appropriate power of the operator $\Phi$
can also be used in the interaction term (10) so as to make the
coupling
constant dimensionless in all the relevant space-time dimensions.

The full action of our space-time scale invariant superstring
theory is
given by
$$S_{superstring} = S_{0} + S_{1}\  .\eqno{(31)}$$
At this stage the two couplings in this expression are independent.
But when coupled to supersymmetric gauge theories as in (10), it is
expected that quantum loop calculations relate these couplings to
that of the gauge theory. The equation resulting from the variation
of the action (31) with respect to $h_{ab}$ leads to the usual
vanishing of the components of the energy-momentum tensor:
$$\Xi_{ab} - {1\over 2}h_{ab}h^{cd}~\Xi_{cd} = 0\eqno{(32)}$$
$$\Xi_{ab} = {1\over 4\pi}(\kappa_{0}^2\tau_{a}\tau_{b} + 
\kappa_{1}^2 G_{ab}\Phi)$$
$$\tau_{a} = {1\over 2} v_{a}^{\mu}\sigma_{\mu}. $$

Let us now consider possible applications of the above formalism.
One
possibility to which we have alluded above is to add the space-time
scale invariant action (31) and the long range interaction (10) to
one of the existing superstring actions and study the resulting
theory to see if it has a (super) QCD-like phase. One would expect
that this theory would be more appropriate for such an analysis
than its bosonic
version. The other possibility is to consider the action (31) as a
superstring theory in its own right. In that case, we note that the
structure of the action (30) suggests an interesting possibility.
Aside from the composite operator $\Phi$, it is identical to the
Green-Schwarz action [9]. If due to the unknown dynamics at the
Planck scale this four fermion operator acquires a non-vanishing
expectation
value, i.e., if   
$${\kappa_{1}^2\over 4\pi}<\Phi> =
\mu_{string~tension}\eqno{(31)}$$
then we have a dynamical basis for understanding how a dimensional
parameter arises in superstring theories. Clearly, much remains to
be done.

\bigskip
This work was supported, in part, by the department of energy under
the
contract number DOE-FG02-84ER40153.

\bigskip
\noindent{\bf References}

\begin{enumerate}

\item {[1]} A. Polyakov, Nucl. Phys. B268 (1986) 406, also 
Gauge fields, and Strings, Harwood academic publishers, 1987 
\item {[2]} M. Awada and D. Zoller, Phys.Lett B325 (1994) 115 ;
M. Awada, Phys. Lett. B351 (1995) 463 ; ibid 468
\item {[3]} M. Kalb, and P. Ramond Phys. Rev. D Vol.9 (1974) 2237
\item {[4]} L.N.Chang, and F. Mansouri, proceeding of the John
Hopkins
workshop, ed. G.Domokos and S.Kovesi Domokos, John Hopkins Univ.
(1974).
\item {[5]} M. Awada and F. Mansouri, in preparation
\item {[6]} M. Awada and F. Mansouri, UCTP-103-1995-hep-th/9512098,
Phys. Lett. B, in press.
\item {[7]} J. Wess and J. Bagger, Introduction to supersymmetry,
Princeton University Press 1983; S.J. Gates, Jr., et al,
Superspace, Benjamin, 1983.
\item {[8]} M. Awada and F. Mansouri, in preparation
\item {[9]} M.Green, J.Schwarz, and E.Witten, Superstring Theory,
Cambridge University press 1987.
\end{enumerate}

\end{document}